\documentclass[
reprint,
preprintnumbers,
 amsmath,amssymb,
 aps,
 prl
]{revtex4-1}

\usepackage{graphicx}
\usepackage{dcolumn}
\usepackage{bm}

\newcommand{\sh}{\hat{s}}
\newcommand{\fs}{\mathcal{F}}

\begin{document}

\preprint{NORDITA 2019-077}

\title{Joule-Thomson Cooling in Graphene}

\author{K. Zarembo}
\affiliation{
 Nordita, KTH Royal Institute of Technology and Stockholm University,
Roslagstullsbacken 23, SE-106 91 Stockholm, Sweden
}%
\affiliation{
 Niels Bohr Institute, Copenhagen University, Blegdamsvej 17, 2100 Copenhagen, Denmark
}

\begin{abstract}
Electrons in graphene exhibit hydrodynamic behavior in a certain range of temperatures. We indicate that in this regime electric current can result in cooling of electron fluid due to the Joule-Thomson effect. Cooling occurs in the Fermi-liquid regime, while for the Dirac fluid the effect results in heating.
\end{abstract}

\maketitle


If collective effects of electron interactions in a very clean sample prevail over impurity scattering, electrons may flow as a viscous fluid \cite{gurzhi1963minimum}. The hydrodynamic regime of electron transport is observed in graphene in a certain range of temperatures \cite{Torre2015nonlocal,LevitovFalkovich2016negative,Bandurin2016negative,Crossno2016WFranz,KrishnaKumar2017superballistic,Bandurin2018fluidity}, and in other 2d materials \cite{Moll2016PdCoO2,Gooth2017WP2}.  Spectacular consequences of hydrodynamic transport include negative local resistivity \cite{Torre2015nonlocal,LevitovFalkovich2016negative,Bandurin2016negative,Bandurin2018fluidity}, violation of ballistic bound on conductance \cite{Torre2015nonlocal,Guo2017higher,KrishnaKumar2017superballistic,Gooth2017WP2} (Gurzhi effect \cite{gurzhi1963minimum}), breakdown of Wiedemann-Franz law \cite{Crossno2016WFranz,Lucas2016hydro,Gooth2017WP2}, and negative magnetoresistance \cite{Lucas2016negative,Gooth2017WP2} (see \cite{LucasChunFong2018review,Narozhny2017review,Narozhny_2019} for reviews). Another possible manifestation of collective electron flow is cooling of electron fluid passing through a narrow constriction. 

In the solid state setting electric current normally generates heat, and
cooling of electron flow may look counterintuitive, but in fluid mechanics this phenomenon is well known and underlies a widely used method of gas refrigeration by throttling, which exploits the Joule-Thomson (JT) effect \cite{thomson1853xiv}. Our goal is to theoretically study the counterpart of the JT effect in graphene. More conventional cooling mechanism due to tunneling through graphene-insulator-superconductor junction has been considered in \cite{Vischi2019cooling}. 

\begin{figure}
\begin{center}
 \centerline{\includegraphics[width=4.5cm]{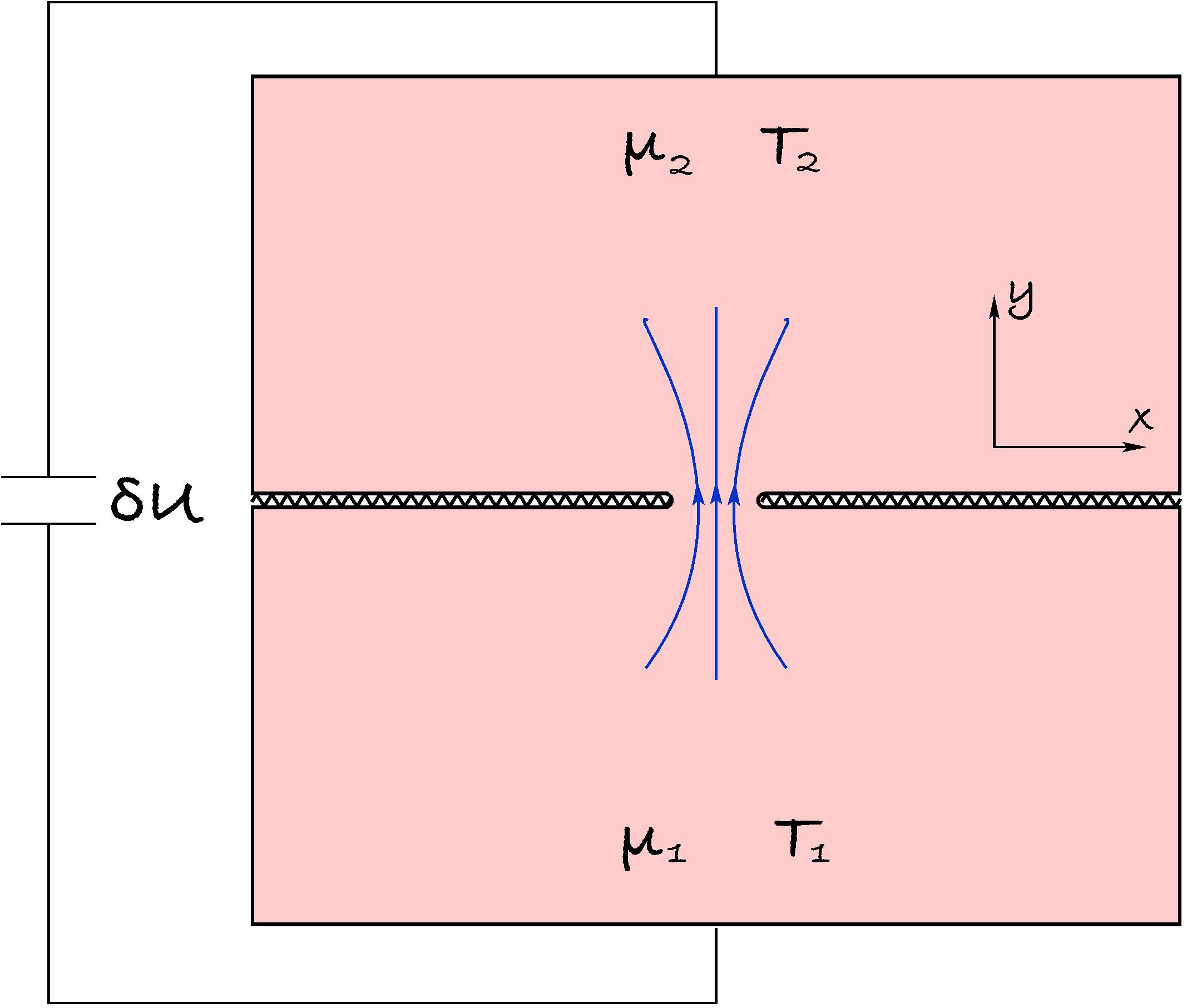}}
\caption{\label{flow}\small Flow through constriction.}
\end{center}
\end{figure}

The simplest realization of throttling is a flow through constriction, illustrated in fig.~\ref{flow}. Consider two strips of graphene connected by a narrow bridge and subject to a constant voltage $\delta U$. Upon passing from one strip to another the temperature of electron fluid may change as a result of work done by external force. Assuming that one strip is kept at temperature $T_1$ and denoting electron temperature in the other by $T_2$, the cooling/heating effect can be characterized by the temperature drop $\delta T=T_1-T_2$ relative to the potential difference $\delta \mu =\mu _1-\mu _2=e\delta U$:
\begin{equation}
 \delta T=\alpha \delta \mu,
\end{equation}
where $\mu _1$ and $\mu _2$ are chemical potentials on the two sides of the bridge.
The dimensionless coefficient $\alpha $ can take either sign and is defined such that $\alpha >0$ corresponds to cooling.

Textbook derivations of the JT effect start with the enthalpy conservation:
\begin{equation}
 \delta\, \frac{\epsilon +P}{n}=0,
\end{equation}
where $\epsilon $ and $P$ are energy density and pressure of the electron fluid and $n$ is the charge carrier density. Thermodynamic relation $\epsilon +P=\mu n+Ts$, where $s$ is the entropy density, then yields
\begin{equation}\label{alpha-gen}
 \alpha =-\frac{A+T\,\frac{\partial \sh}{\partial \mu }}{A\sh+T\,\frac{\partial \sh}{\partial T}}\,,
\end{equation}
where $\sh=s/n$ is specific entropy and $A=1$. We shall later see that effects of viscous heating and momentum dissipation can be absorbed into redefinition of $A$ while preserving the same functional form of the cooling coefficient. For the time being we keep $A$ as a parameter.

Thermodynamically electrons in graphene behave as 2d Fermi gas with linear dispersion relation, and their pressure is given by
\begin{equation}
 P=4T\int_{}^{}\frac{d^2p}{(2\pi \hbar)^2}\sum_{q =\pm}^{}
 \ln\left(1+\,{\rm e}\,^{\frac{q\mu  -v_F|\mathbf{p}|}{T}}\right),
\end{equation}
where $v_F$ is the Fermi velocity, $q$ labels particles/holes, and the overall factor of four takes into account valley and spin degeneracy.  We tacitly assume that holes and electrons are in thermodynamic equilibrium, which is not a good approximation in neutral graphene. Near the neutrality point electrons and holes should be treated separately \cite{FosterAleiner,Narozhny_2019}, so our derivation will break down near $\mu =0$.

The rest of thermodynamic quantities can be calculated from $dP=nd\mu +sdT$. When applied to (\ref{alpha-gen}) the standard thermodynamic machinery gives
\begin{equation}\label{1over}
 \frac{1}{\alpha }=\frac{3A\,\frac{\fs}{\fs'}}{A+2-3\,\frac{\fs\fs''}{{\fs'}^2}}-\xi,
 \qquad \xi =\frac{\mu }{T}\,,
\end{equation}
where
\begin{equation}
 \fs(\xi )=\mathop{\mathrm{Li}}\nolimits_3(-\,{\rm e}\,^{\xi })
 +\mathop{\mathrm{Li}}\nolimits_3(-\,{\rm e}\,^{-\xi }),
 \end{equation}
and $\mathop{\mathrm{Li}}\nolimits_3$ is the polylogarithm function.

\begin{figure}[t]
\begin{center}
 \centerline{\includegraphics[width=5cm]{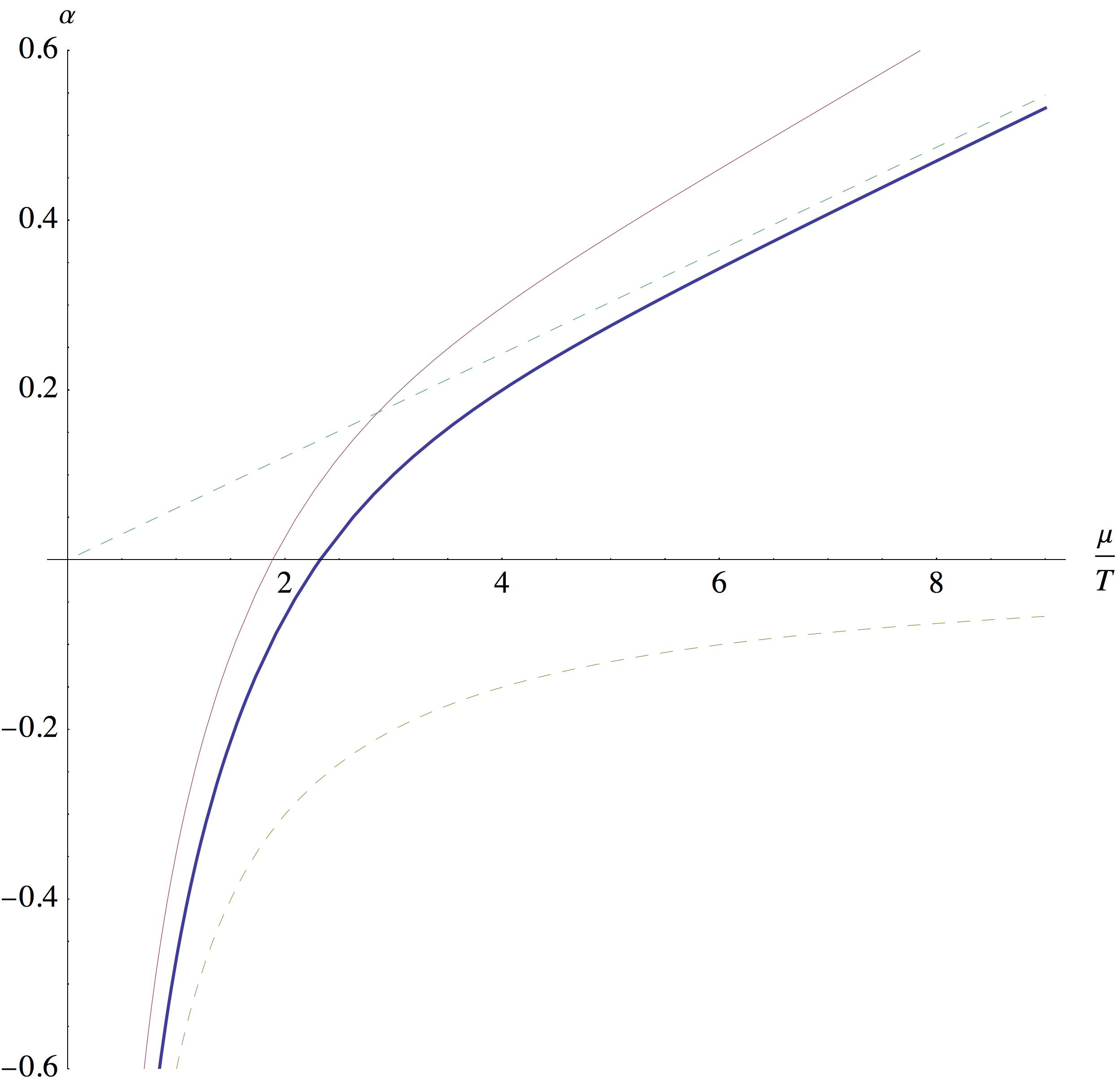}}
\caption{\label{JT}\small The JT coefficient (\ref{1over}) for $A=2/3$. The thin line is the ideal thermodynamics result ($A=1$), shown for comparison. The dashed lines are approximations (\ref{Diracfl}) and (\ref{Fermilq}). }
\end{center}
\end{figure}

Expanding the general formula at small or large $\xi $, we find that in the Fermi liquid regime ($\mu \gg T$) the JT effect results in cooling:
\begin{equation}\label{Fermilq}
 \alpha \simeq \frac{3A\mu }{2(1+A)\pi ^2T}\qquad (\mu \gg T),
\end{equation}
as long as $A>0$ (recall that thermodynamic relations give $A=1$). At the same time, in the Dirac fluid regime ($\mu \ll T$) the JT coefficient is negative which corresponds to heating:
\begin{equation}\label{Diracfl}
 \alpha \simeq -\frac{T}{(1+A)\mu }\qquad (\mu \ll T).
\end{equation}
The effect changes sign at  the inversion point.  The existence of such is well familiar from thermodynamics of ordinary gases. The JT coefficient for the full range of chemical potentials is plotted in fig.~\ref{JT}. At the neutrality point the coefficient is infinite signaling the breakdown of our assumptions. As explained above the failure of the derivation has to do with absence of local equilibrium between electrons and holes at the neutrality point. 

The  thermodynamic derivation of the JT effect relies on enthalpy conservation, but enthalpy production in the moving electron fluid may be substantial. In general, mechanical effects of the flow cannot be neglected. We will not attempt to 
develop comprehensive theory, but will consider two effects not accounted for by simple thermodynamics, the viscous heating by mechanical shear and Ohmic resistance due to momentum relaxation. As we shall see, viscous heating is the prime source of enthalpy production and leads to order one effects, while effects of Ohmic resistance are parametrically small with the size of opening in the constriction.

In the hydrodynamic regime a stationary electron flow is described by the Navier-Stokes equations \cite{Lucas2016hydro}:
\begin{eqnarray}\label{ns}
 &&\partial _i(nv^i)=0
\nonumber \\
&&\partial _i\left\{\left[(\epsilon +P)\delta ^{ij}-\eta \Pi ^{ij}\right]v_j\right\}=0
\nonumber \\
&&\partial ^jP-\eta \partial _i\Pi ^{ij}=0,
\end{eqnarray}
where $\Pi _{ij}=\partial _iv_j+\partial _jv_i-\delta _{ij}\partial _kv^k$ is the shear tensor and $\eta $ is the shear viscosity. These equations constitute four conditions for four unknowns -- the local temperature, the chemical potential, and the two components of velocity. The bulk viscosity $\zeta \delta _{ij}\partial _kv^k$ vanishes in graphene due to underlying conformal invariance and, in addition, we only consider divergence-free flows for which the bulk viscosity does not contribute. We also assume that electrons are locked to holes or that holes are absent due to charge imbalance. Then motion of the electron fluid can then be described as a flow of charge, otherwise one needs to introduce separate hydrodynamic variables for holes and electrons \cite{FosterAleiner}.  

The first two equations combined give
\begin{equation}\label{enthalpy}
 v^i\partial _i\,\frac{\epsilon +P}{n}=\frac{\eta }{n}\,\partial _i\left(\Pi ^{ij}v_j\right).
\end{equation}
In absence of viscosity, specific enthalpy is constant along the flow lines, as was assumed in the simple-minded thermodynamic derivation. Non-linear terms of order $v/v_F$, neglected because $v\ll v_F$, presumably become important for very small viscosity, but we assume that viscosity is sufficiently large to suppress non-linearities. Notice that viscosity can be absorbed into rescaling $v\rightarrow \eta ^{-1}v$, and will drop from the final answer for the JT coefficient. 

It is instructive to rewrite enthalpy non-conservation in a different form. Combining (\ref{enthalpy}) with the third equation in (\ref{ns}) we get:
\begin{equation}\label{eproduction}
 v^i\partial _i\sh=\frac{\eta }{2nT}\,\Pi ^{ij}\Pi _{ij}.
\end{equation}
The right-hand side is the manifestly positive entropy production rate due to the viscous shear. It may be expected on general grounds that viscosity leads to stronger heating or weaker cooling and thus diminishes the JT coefficient compared to the thermostatic consideration. We are now going to quantify this effect.

The only assumptions so far were stationarity of the flow and $v\ll v_F$. If in addition we assume that electrons move slowly compared to any other scale, the equations simplify and can be essentially linearized. For instance, assuming that all gradients are small, $\partial n v\ll n\partial v$, the first two equations in (\ref{ns}) become equivalent to the divergence free condition:
\begin{eqnarray}\label{simplified}
 &&\partial _iv^i=0
\nonumber \\
&&\eta \partial ^2v_i=\partial _iP.
\end{eqnarray}
These are Stokes equations for a stationary flow of an incompressible fluid at low Reynolds number. They can  
be solved exactly for the geometry in fig.~\ref{flow} \cite{Guo2017higher}:
\begin{eqnarray}\label{finitea-solution}
 P&=&P_0-\frac{4\eta u}{a}\,\mathop{\mathrm{Im}}\frac{z}{\sqrt{\frac{a^2}{4}-z^2}}
\nonumber \\
v_x&=&\frac{2uy}{a}\,\mathop{\mathrm{Re}}\frac{z}{\sqrt{\frac{a^2}{4}-z^2}}
\nonumber \\
v_y&=&\frac{2u}{a}\,\mathop{\mathrm{Re}}\sqrt{\frac{a^2}{4}-z^2}
-\frac{2uy}{a}\,\mathop{\mathrm{Im}}\frac{z}{\sqrt{\frac{a^2}{4}-z^2}}\,.
\end{eqnarray}
Here $z=x+iy$, $a$ is the width of the bridge, and $u$ is the maximal  velocity attained by the fluid. The square root is analytic on the complex plane with a semi-infinite cut representing the constriction. 

The solution obeys the no-slip boundary conditions: $v_i(x,0)=0$ for $|x|>a$. Another commonly discussed possibility is no-stress, $v_y=0$,  
$\partial _yv_x=0$. While experimental results are consistent with no-slip  \cite{Moll2016PdCoO2,Gooth2017WP2}, theory arguments suggest that the no-stress conditions are more realistic at low temperatures \cite{KiselevSchmalian}. It would be interesting to repeat the calculation for this case or for more general mixed boundary conditions  \cite{KiselevSchmalian}.

The flow (\ref{finitea-solution}) is sustained by the pressure drop:
\begin{equation}
 \delta P=\frac{8\eta u}{a}\,.
\end{equation}
The loss function can be readily calculated for this solution:
\begin{equation}
 \Pi ^{ij}\Pi _{ij}=\frac{2u^2a^2y^2}{\left|\frac{a^2}{4}-z^2\right|^3}\,.
\end{equation}
Integrating the entropy production rate along the midflow according to  (\ref{eproduction}) we get:
\begin{equation}
 -\delta \sh=\frac{2\eta ua }{nT}\int\limits_{-\infty }^{+\infty }
 \frac{dy\,y^2}{\left(\frac{a^2}{4}+y^2\right)^{\frac{5}{2}}}
 =\frac{16\eta u}{3anT}=\frac{2}{3nT}\,\delta P.
\end{equation}
The relation $\delta P=n\delta \mu +s\delta T$ then gives the same formula (\ref{alpha-gen}) for the JT coefficient, but with $A=2/3$. 
Alternatively the same result can be derived by computing the enthalpy production with the help of (\ref{enthalpy}). 

The resulting JT coefficient is displayed in fig.~\ref{JT}. As expected, viscosity diminishes the cooling effect in the whole range of parameters, leading to stronger heating or weaker cooling compared to ideal thermodynamics. Interestingly, all the dependence on the velocity, geometry of the constriction and even on shear viscosity cancels out leaving behind order one reduction in the cooling power.

We now turn to Ohmic resistivity.
One may anticipate, on general grounds, that impurity scattering and interaction with phonons have a smaller thermal effect than viscosity for slow, low Reynolds number flows. The argument essentially follows from dimensional analysis.
Momentum non-conservation due to umklapp or impurity scattering at a rate $\tau _{\rm imp}^{-1}$ is characterized macroscopically by a length scale \cite{LucasChunFong2018review}:
\begin{equation}
 \lambda =v_F\sqrt{\frac{\eta \tau _{\rm imp}}{\epsilon +P}}\,.
\end{equation}
The momentum-relaxation length $\lambda $ is estimated to lie between a fraction \cite{Torre2015nonlocal,Bandurin2016negative} to a few \cite{KrishnaKumar2017superballistic} microns. We assume that the opening in the constriction is smaller: $a\ll \lambda $. In this regime the flow is affected by momentum relaxation only far away from the constriction, for $|z|\sim \lambda $, while the two factors important for the JT effect, the pressure drop and the entropy production, mostly occur at $|z|\sim a$. The ensuing corrections to the JT coefficient are thus small, suppressed by the ratio $a/\lambda $. A more careful analysis shows that the effect is logarithmically enhanced if the total size of the system is much larger than the momentum-relaxation length: $L\gg \lambda $.  

Hydrodynamically, momentum relaxation is described by an additional damping term in the  Navier-Stokes equation \cite{LucasChunFong2018review}:
\begin{equation}
 \partial ^jP-\eta \partial _i\Pi ^{ij}=-\frac{\eta }{\lambda ^2}\,v^j,
\end{equation}
which, in turn, contributes to the entropy production:
\begin{equation}
 v^i\partial _i\sh=\frac{\eta }{nT}\left(\frac{1}{\lambda ^2}\,v^iv_i+\frac{1}{2}\,\Pi ^{ij}\Pi _{ij}\right),
\end{equation}
and also changes the last equation in (\ref{simplified}):
\begin{equation}
 \eta \partial ^2v_i-\frac{\eta }{\lambda ^2}\,v_i=\partial _iP.
\end{equation}

The flow equations for generic $a$ and $\lambda $ can only be solved numerically, but for $a\ll \lambda $ an approximate analytic solution can be constructed. The near zone is accurately described by (\ref{finitea-solution}) so long as $|z|\ll \lambda $, while for $|z|\gg a$ the opening in the constriction can be approximated by the delta-function. In that approximation the solution can again be found analytically, generalizing the known result at $\lambda =\infty $ \cite{Falkovich2017linking} to finite $\lambda $. The strategy is to Fourier transform in $x$ and solve the ensuing linear ODEs with the boundary conditions
\begin{equation}
 v_x(k;0)=0,\qquad v_y(k;0)=\frac{\pi ua}{4}\,.
\end{equation}
 Straightforward algebraic manipulations give
\begin{eqnarray}
v_x&=&\frac{ua\lambda ^2}{4}\int\limits_{0}^{\infty }dk\,
\omega\left(\omega +k \right)
\left(\,{\rm e}\,^{-ky}-\,{\rm e}\,^{-\omega y}\right)
\sin kx
\nonumber \\
v_y&=&\frac{ua\lambda ^2}{4}\int\limits_{0}^{\infty }dk\,
\left(\omega +k\right)
\left(\omega \,{\rm e}\,^{-ky}-k\,{\rm e}\,^{-\omega y}\right)
\cos kx
\nonumber \\
P&=&\bar{P}_0+\frac{\eta ua}{4}\int\limits_{0}^{\infty }\frac{dk}{k}\,\,
\omega (\omega +k)\cos kx\,{\rm e}\,^{-ky}
\nonumber \\
\omega &=&\sqrt{k^2+\frac{1}{\lambda ^2}}\,.
\end{eqnarray}
The small-$k$ divergence of the integral defining pressure is actually fake, it can be absorbed into the constant of integration, but the ensuing log behavior at large distances will have interesting bearing on the JT effect. The integral can be  explicitly evaluated:
\begin{equation}
 P=P_0-\frac{4\eta u}{a}-\frac{\eta ua}{4\lambda ^2}\,\mathop{\mathrm{Re}}G\left(\frac{z}{i\lambda }\right),
\end{equation}
 where the constant of integration is fixed by matching to the near-zone solution. The function $G(s)$ is given by
\begin{equation} 
 G(s)=\ln s-\frac{1}{s^2}+\frac{\pi }{2s}\left(Y_1(s)-\mathbf{H}_1(s)\right),
\end{equation}
where $Y_1$, $\mathbf{H}_1$ are Neumann and Struve functions, respectively. The solution is valid for $y>0$. One can readily check that the near and far zone solutions match in their overlapping region of validity $a\ll |z|\ll \lambda $.

In the opposite regime of $|z|\gg \lambda $ the pressure grows logarithmically. For instance, at $x=0$ and $y\gg \lambda $,
\begin{equation}
 P\simeq P_0-\frac{4\eta u}{a}-\frac{\eta ua}{4\lambda ^2}\,\ln\frac{y}{\lambda }\,,\qquad v_y\simeq \frac{ua}{4y}\,.
\end{equation}
The velocity also drops too slowly, and the total entropy production will log diverge. These results are easy to understand qualitatively. When propelling electrons through the sample, the external force must do extra work to counter momentum relaxation. This leads to a larger pressure drop. The entropy production increases due to additional Ohmic losses. Both effects turn out to be logarithmically enhanced.

 The total pressure drop is thus defined by the size of the whole sample:
\begin{equation}
 \delta P=\frac{8\eta u}{a}\left(1+\frac{a^2}{16\lambda^2 }\,\ln\frac{L}{\lambda }\right).
\end{equation}
Likewise, the total entropy production is log-divergent:
\begin{equation}
 -\delta \sh=\frac{16\eta u}{3anT}\left(1+\frac{3a^2}{32\lambda ^2}\,\ln\frac{L}{\lambda }\right).
\end{equation}
The JT coefficient then takes the form (\ref{alpha-gen}) with 
\begin{equation}
 A=\frac{2}{3}\left(1+\frac{a^2}{32\lambda ^2}\ln\frac{L}{\lambda }\right),
\end{equation}
in the logarithmic approximation.
The correction due to momentum relaxation is parametrically small with $a$, but the coefficient is logarithmically enhanced for a large sample. For a more realistic situation of $L\sim \lambda $ the correction should still be of order $a^2/\lambda ^2$, but the coefficient is more difficult to calculate and will certainly depend on geometry.

Hydrodynamic nature of electron flow in biased graphene leads to JT cooling  when the current is forced through a narrow constriction. Cooling occurs in the Fermi liquid regime, for sufficiently large charge imbalance or at sufficiently low temperatures. In the geometric setting at hand the inversion point lies at $\mu_{\rm inv} =3.32T$ (fig.~\ref{JT}). For lower chemical potentials the JT effect results in heating, which is most pronounced in the Dirac liquid regime of $\mu \ll T$. Although similar to conventional Joule heating the mechanics behind this effect is quite different, in particular the temperature increment is linear in applied voltage and not quadratic. 
We expect, but have not checked, that effects of the conventional Joule heating due to  normal resistivity are parametrically smaller than viscosity-driven entropy production.
It is necessary to keep in mind however that our derivation is invalid close to the  neutrality point.

It would be interesting to study the JT effect for different geometries of the flow and for more general boundary conditions. Our assumptions break down at the neutrality point where electrons and holes are not in equilibrium with each other and to develop reliable theory for neutral graphene one would need to treat the flow of electrons and holes separately \cite{FosterAleiner}. Finally, it would be interesting to understand the strong-coupling regime, perhaps with the help of the AdS/CFT methods \cite{Erdmenger:2018svl}.

\begin{acknowledgements}
The author would like to thank A. Balatsky and G. Falkovich for  discussions. This work  was supported by the Swedish Research Council (VR) grant
2013-4329, by the grant "Exact Results in Gauge and String Theories" from the Knut and Alice Wallenberg foundation, and by RFBR grant 18-01-00460A. 
\end{acknowledgements}


\end{document}